\documentclass[aps,PRB,twocolumn,superscriptaddress]{revtex4-2}
\usepackage[T1]{fontenc}
\usepackage{times}
\usepackage{graphicx,color}
\usepackage{amsfonts,amsmath,amssymb,amsbsy}
\usepackage[colorlinks=true, linkcolor=blue, citecolor=blue, urlcolor=blue]{hyperref}
\usepackage{float}

\begin{document}

\title{Almost half-quantized  planar Hall effects in $X$-wave magnets
with $X=p,d,f,g,i$}
\author{Motohiko Ezawa}
\affiliation{Department of Applied Physics, The University of Tokyo, 7-3-1 Hongo, Tokyo
113-8656, Japan}

\begin{abstract}
The planar Hall effect is a phenomenon that the Hall conductivity emerges
perpendicular to the electric field in the presence of an in-plane magnetic
field. We investigate the planar Hall effect in two-dimensional  metal
coupled with  higher symmetric $X$-wave magnets with $X=p,d,f,g,i$,\
where those with $X=d,g,i$ are altermagnets. The $X$-wave magnet is
characterized by the number $N_{X}$ of the nodes in the band structure,
where $N_{X}=1,2,3,4,6$ corresponding to $X=p,d,f,g,i$.  Although the
system is metallic, provided the Dirac gap is tiny, we  demonstrate that
the Hall conductivities are almost half quantized and well approximated by
the formula $\sigma _{xy}=\pm (e^{2}/2h)$ sgn$\left( J\sin N_{X}\Phi \right) 
$, where $J$ is the coefficient of the coupling between the $X$-wave magnet
and the electrons, and $\Phi $ is the direction of the applied magnetic
field. Hence, the Hall conductivity is periodic in $\Phi $, and the
periodicity is equal to the number $N_{X}$ of the nodes. This property may
be used to confirm that the target material is indeed an $X $-wave magnet.
Furthermore, the sign of $J$ may be used as a bit for antiferromagnetic
spintronics.
\end{abstract}

\date{\today }
\maketitle

\section{Introduction}

\begin{figure}[t]
\centerline{\includegraphics[width=0.48\textwidth]{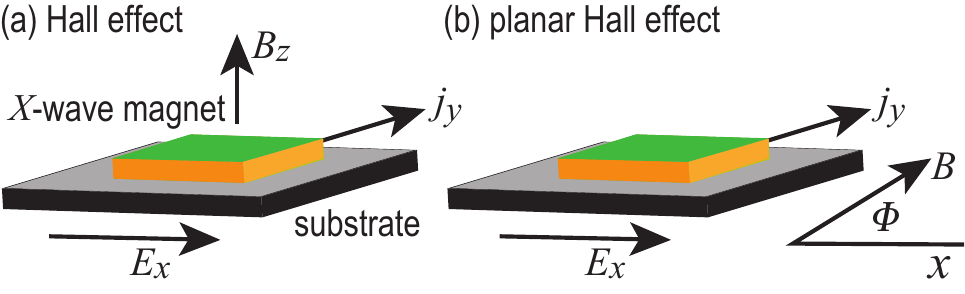}}
\caption{Illustration for (a) Hall effect and (b) planar Hall effect. In the
planar Hall effect, when the electric field is applied along the $x$ axis
and the magnetic field $(B\cos\Phi,B\sin\Phi ,0)$ is applied parallel to the
system, the Hall current flows along the $y$ axis. The Hall conductivity is
predicted to be given by the formula (\protect\ref{pHall}). }
\label{FigIllust}
\end{figure}

The Hall effect is a prominent phenomenon in two-dimensional materials. A
current flows into a direction perpendicular to the applied electric field,
when the magnetic field is applied perpendicular to the plane, as
illustrated in Fig.\ref{FigIllust}(a). Similarly, a current flows into a
direction perpendicular to the applied electric field, when the magnetic
field is applied parallel to the plane, as illustrated in Fig.\ref{FigIllust}%
(b). It is the planar Hall effect\cite%
{Tang,XLiu,Nandy,Taskin,Burkov,Kumar,Zliu}. We are interested in the planar
Hall effect in altermagnets and related materials.

An altermagnet is expected to be a promising candidate for antiferromagnetic
spintronics\cite{SmejX,SmejX2}, where the direction of the N\'{e}el vector
is used as a bit. Antiferromagnets are thought to be ideal for future
spintronics memory with ultra-fast and ultra-dense application due to the
lack of the stray field in contrast to ferromagnets\cite%
{Jung,Baltz,Han,Ni,Godin,Kimura,ZhangNeel}. However, it is hard to detect
the direction of the N\'{e}el vector because net magnetization is zero in
antiferromagnets. Nevertheless, this is possible by measuring anomalous Hall
conductivity in altermagnets because the altermagnets break time-reversal
symmetry\cite{Fak,Tsch,Sato,Leiv}. Indeed, it is theoretically predicted\cite%
{Fak,Sato,Attias,Sheoran}, and furthermore experimentally observed\cite%
{Feng,Tsch,Leiv,Seki,Reich}. Very recently, the planar Hall effect is
discussed in $d$-wave altermagnets\cite{Korra,BFChen},  where it is found
to be almost half quantized\cite{Korra}.

In general, it is interesting to study the higher symmetric $X$-wave magnets%
\cite{SmejX,SmejX2,pwave,GI} with $X=p,d,f,g,i$. They are characterized by
the specific band structure, where the number of the nodes is $%
N_{X}=1,2,3,4,6$, respectively. The $X$-wave magnets with $X=d,g,i$ are
altermagnets. The effective Hamiltonian\cite{GI} for the $X$-wave magnet is
concisely described by a two-band model containing the term $J\sigma
_{z}f_{X}\left( \mathbf{k}\right) $, where $J$ is the coefficient of the
coupling between the $X$-wave magnet and the electrons, $\sigma _{z}$ is the
Pauli matrix representing the spin of electrons, and $f_{X}\left( \mathbf{k}%
\right) $\ is a symmetric function characterizing the $X$-wave magnet.

In this paper, we show that Hall conductivities emerge in two-dimensional 

metallic systems coupled with  various $X$-wave magnets provided magnetic
field is applied along the in-plane direction. It is induced by the momentum
shift of the Dirac cone accompanied by the gap opening. The main result is
that the Hall conductivity is  well approximated  by the  %
half-quantization  formula%
\begin{equation}
\sigma _{xy}=\frac{e^{2}}{h}\frac{\left( -1\right) ^{s_{X}}}{2}\text{sgn}%
\left( J\frac{B^{N_{X}}}{\lambda ^{N_{X}}}\sin N_{X}\Phi \right) ,
\label{pHall}
\end{equation}%
where $s_{X}=1,1,-1,-1,1$ for $X=p,d,f,g,i$, respectively. Here $B>0$ is the
applied magnetic field, $\Phi $ is the direction of the applied magnetic
field, and $\lambda $ is the coefficient of the spin-orbit interaction.  %
It is curious that the Hall conductivity is almost half quantized although
the system is metallic. This occurs when the Berry curvature is almost
localized in the vicinity of the Dirac point with a tiny Dirac gap.  The
Hall conductivity is proportional to the sign of $J$. Hence, it is possible
to use the sign of $J$\ as a bit, while it is possible to detect the sign of 
$J$\ by observing the planar Hall effect. In addition, the number of the
nodes $N_{X}$\ is detectable by measuring the periodicity of the Hall
conductivity as a function of $\Phi $.

\section{Higher symmetric $X$-wave magnets}

We analyze a  metallic  system where an $X$-wave magnet is placed on a
substrate as in Fig.\ref{FigIllust}. The minimal two-band Hamiltonian is
given by%
\begin{equation}
H\left( \mathbf{k}\right) =\frac{\hbar ^{2}\mathbf{k}^{2}}{2m}+\lambda
\left( \mathbf{k}\times \mathbf{\sigma }\right) _{z}+Jf_{X}\left( \mathbf{k}%
\right) \sigma _{z}+\mathbf{B}\cdot \mathbf{\sigma }  \label{2dHamil}
\end{equation}%
in two dimensions with the use of the Pauli matrix $\mathbf{\sigma }$. The
first term represents the kinetic energy,  making the system metallic, 

the second term the Rashba interaction introduced by making an interface
between the $X$-wave magnet and the substrate, the third term the $X$-wave
term with the $X$-wave function $f_{X}\left( \mathbf{k}\right) $, and the
last term the applied magnetic field $\mathbf{B}=(B_{x},B_{y},0)$. The
Rashba interaction forms a Dirac cone at the $\Gamma $\ point. The Rashba
interaction is introduced by placing an altermagnet on the substrate\cite%
{SmejRev,SmejX,SmejX2,Zu2023,Gho,Li2023,EzawaAlter,EzawaMetricC}.

The $X$-wave function $f_{X}\left( \mathbf{k}\right) $ reads as follows\cite%
{SmejX,SmejX2,pwave,GI,Fukaya},%
\begin{align}
f_{s}& =1,  \label{EqX} \\
f_{p}& =k_{x}=k\cos \phi ,  \label{p-f} \\
f_{d}& =2k_{x}k_{y}=k^{2}\sin 2\phi ,  \label{d-f} \\
f_{f}& =k_{x}\left( k_{x}^{2}-3k_{y}^{2}\right) =k^{3}\cos 3\phi ,
\label{f-f} \\
f_{g}& =4k_{x}k_{y}\left( k_{x}^{2}-k_{y}^{2}\right) =k^{4}\sin 4\phi ,
\label{g-f} \\
f_{i}& =2k_{x}k_{y}\left( 3k_{x}^{2}-k_{y}^{2}\right) \left(
k_{x}^{2}-3k_{y}^{2}\right) =k^{6}\sin 6\phi ,  \label{i-f}
\end{align}%
where $k_{x}=k\cos \phi $, $k_{y}=k\sin \phi $.

\section{Hall conductivity}

 We review the Hall conductivity in the  two-band Hamiltonian in the
form of 
\begin{equation}
H=h_{0}+\sum_{j=x,y,z}h_{j}\sigma _{j}.
\end{equation}%
We define a normalized vector%
\begin{equation}
\mathbf{n}\left( \mathbf{k}\right) \equiv \mathbf{h}/|\mathbf{h}|=\left(
\sin \theta \cos \phi ,\sin \theta \sin \phi ,\cos \theta \right) .
\end{equation}%
The Berry curvature is expressed in terms of the solid angle of the vector $%
\mathbf{n}$ as\cite{Hsiang,Stic,Jiang}%
\begin{equation}
\Omega _{\pm }\left( \mathbf{k}\right) =\pm \frac{1}{2}\mathbf{n}\cdot
\left( \partial _{k_{x}}\mathbf{n}\times \partial _{k_{y}}\mathbf{n}\right) ,
\label{Berry}
\end{equation}%
for the upper band ($+$) and the lower band ($-$). The integration of the
Berry curvature over whole Brillouin zone gives the Chern number $C_{\pm }$,%
\begin{equation}
\frac{1}{2\pi }\int_{\text{BZ}}\Omega _{\pm }\left( \mathbf{k}\right) d%
\mathbf{k}=C_{\pm }.
\end{equation}%
It is interpreted as the wrapping number of the vector $\mathbf{n}$ known as
the Pontryagin number.

According to the Thouless-Kohmoto-Nightingale-den Nijs\ formula, the Hall
conductivity is related to the Berry curvature as 
\begin{equation}
\sigma _{xy}=\frac{e^{2}}{2\pi h}\sum_{n=\pm }\int d^{2}k\,f\left(
E_{n}\left( \mathbf{k}\right) \right) \Omega _{n}\left( \mathbf{k}\right) ,
\label{Hall}
\end{equation}%
where $f\left( E_{\pm }\left( \mathbf{k}\right) \right) =1/[\exp \frac{%
E_{\pm }(\mathbf{k})-\mu }{k_{\text{B}}T}+1]$ is the Fermi distribution
function, and $\mu $ is the chemical potential. In an insulator at zero
temperature, the Hall conductivity is proportional to the Chern number $C_{-}
$ of the bottom band,%
\begin{equation}
\sigma _{xy}=\frac{e^{2}}{h}C_{-}.
\end{equation}%
In the Dirac system, the Chern number is half quantized if the chemical
potential is within the Dirac gap. Hence, the Hall conductivity per Dirac
cone is given by%
\begin{equation}
\sigma _{xy}=\pm \frac{e^{2}}{2h}.
\end{equation}%
This quantization formula is exact only for insulators.

\section{Planar Hall effects}

\label{SecPlanarHE}

 The above argument is not applicable to a metallic system in general.
However, it is possible to define the Berry curvature to the two-band
Hamiltonian (\ref{2dHamil}) provided two bands are gapped and non-degenerate.%

The Dirac point shifts in the presence of the in-plane magnetic field. The
shifted Dirac point is given by $(k_{x}^{\prime },k_{y}^{\prime })=0$, where%
\begin{equation}
k_{x}^{\prime }=k_{x}+\frac{B_{y}}{\lambda },\qquad k_{y}^{\prime }=k_{y}-%
\frac{B_{x}}{\lambda },
\end{equation}%
by solving the equation%
\begin{equation}
\lambda \left( k_{x}\sigma _{y}-k_{y}\sigma _{x}\right) +B_{x}\sigma
_{x}+B_{y}\sigma _{y}=0.
\end{equation}%
In terms of the shifted momentum, the Hamiltonian (\ref{2dHamil}) is given by%
\begin{equation}
H=H_{\text{kine}}^{\prime }+H_{\text{R}}^{\prime }+H_{\text{mass}}^{\prime },
\end{equation}%
up to the first order of $k_{x}^{\prime }$\ and $k_{y}^{\prime }$, where $H_{%
\text{kine}}^{\prime }$ is the kinetic energy at the shifted Dirac point, 
\begin{equation}
H_{\text{kine}}^{\prime }=\frac{\hbar ^{2}}{2m}\left[ \left( \frac{B}{%
\lambda }\right) ^{2}-\frac{2B_{y}}{\lambda }k_{x}^{\prime }+\frac{2B_{x}}{%
\lambda }k_{y}^{\prime }\right] ,
\end{equation}%
$H_{\text{R}}^{\prime }$ is the Rashba interaction term, 
\begin{equation}
H_{\text{R}}^{\prime }=\lambda \left( k_{x}^{\prime }\sigma
_{y}-k_{y}^{\prime }\sigma _{x}\right) ,
\end{equation}%
and $H_{\text{mass}}^{\prime }$ is the Dirac mass term,%
\begin{equation}
H_{\text{mass}}^{\prime }=J\left( \Delta +g_{x}k_{x}^{\prime
}+g_{y}k_{y}^{\prime }\right) \sigma _{z},
\end{equation}%
with the Dirac gap, 
\begin{equation}
\Delta \equiv f_{X}\left( k_{x}^{\prime }=0,k_{y}^{\prime }=0\right) ,
\label{sDirac}
\end{equation}%
and%
\begin{equation}
g_{x}\equiv \frac{\partial f_{X}\left( k_{x}^{\prime },k_{y}^{\prime
}\right) }{\partial k_{x}^{\prime }},\quad g_{y}\equiv \frac{\partial
f_{X}\left( k_{x}^{\prime },k_{y}^{\prime }\right) }{\partial k_{y}^{\prime }%
}.
\end{equation}%
 The Dirac gap is explicitly determined as%
\begin{equation}
\Delta =\left( -1\right) ^{s_{X}}\frac{B^{N_{X}}}{\lambda ^{N_{X}}}\sin
N_{X}\Phi .  \label{sDirac2}
\end{equation}%
 Then, the Hamiltonian is rewritten as 
\begin{align}
H=& \frac{\hbar ^{2}}{2m}\left[ \left( \frac{B}{\lambda }\right) ^{2}-\frac{%
2B_{y}}{\lambda }k_{x}^{\prime }+\frac{2B_{x}}{\lambda }k_{y}^{\prime }%
\right]   \notag \\
& +\lambda \left( k_{x}^{\prime }\sigma _{y}-k_{y}^{\prime }\sigma
_{x}\right)   \notag \\
& +J\left( \Delta +g_{x}k_{x}^{\prime }+g_{y}k_{y}^{\prime }\right) \sigma
_{z}.
\end{align}

 The Berry curvature (\ref{Berry}) is calculated in the $(k_{x}^{\prime
},k_{y}^{\prime })$ space as 
\begin{equation}
\Omega _{\pm }=\frac{\mp J\Delta \lambda ^{2}}{2\left( \lambda ^{2}k^{\prime
2}+\left( J\Delta +Jg_{x}k_{x}^{\prime }+Jg_{y}k_{y}^{\prime }\right)
^{2}\right) ^{3/2}}.  \label{BerryA}
\end{equation}%
We consider a simple case that the Fermi surface is a circle with the radius 
$k_{\text{c}}$. This is realized for small $J$, where $k_{\text{c}}$ is
calculated as

\begin{equation}
k_{\text{c}}=\sqrt{2m}\sqrt{\mu +m\lambda ^{2}+\sqrt{\Delta ^{2}+2\mu
m\lambda ^{2}+m^{2}\lambda ^{4}}}.
\end{equation}%
We assume that the chemical potential is within the gap. We integrate the
Berry curvature (\ref{BerryA}) with a cutoff $k_{\text{c}}$ for the bottom
band as%
\begin{equation}
\int_{0}^{k_{\text{c}}}kdk\int d\phi \Omega _{-}=-\frac{1}{2}\text{sgn}%
J\Delta +\int d\phi W(k_{\text{c}},\phi ),  \label{JdW}
\end{equation}%
where%
\begin{equation}
W(k_{\text{c}},\phi )\equiv -\frac{J\Delta +k_{\text{c}}g_{\parallel }\left(
\phi \right) }{2\sqrt{\lambda ^{2}k_{\text{c}}^{2}+\left( J\Delta +k_{\text{c%
}}g_{\parallel }\left( \phi \right) \right) ^{2}}}
\end{equation}%
with 
\begin{equation}
g_{\parallel }\left( \phi \right) \equiv g_{x}\cos \phi +g_{y}\sin \phi .
\end{equation}%
The Hall conductivity is given by 
\begin{equation}
\sigma _{xy}=-\frac{e^{2}}{h}\frac{1}{2}\left( 1-\frac{\left\vert J\Delta
\right\vert }{\sqrt{\left( J\Delta \right) ^{2}+\lambda ^{2}k_{\text{c}}^{2}}%
}\right) \text{sgn}J\Delta .  \label{SB}
\end{equation}%
It follows from Eq.(\ref{SB}) that 
\begin{equation}
\left\vert \sigma _{xy}\right\vert <\frac{e^{2}}{h}\frac{1}{2}.
\end{equation}%
The Hall conductivity is not half quantized for a metallic system. However,
for $J\Delta /\lambda k_{\text{c}}\ll 1$, we obtain the half quantization,%
\begin{equation}
\sigma _{xy}=-\frac{e^{2}}{h}\frac{1}{2}\text{sgn}J\Delta ,  \label{HalfQ}
\end{equation}%
which is equivalent to the formula (\ref{pHall}) with the use of Eq.(\ref%
{sDirac2}).

 This phenomenon is understood as follows. The integral of the Berry
curvature over the whole Brillouin zone gives an exactly half-quantized
value if the chemical potential is within the gap for the insulating system.
However, the present system is metallic due to the presence of the kinetic
term. Then, the integration is done only within the Fermi surface. It gives
a non half-quantized Hall conductivity, which is smaller than the
half-quantized value as in Eq.(\ref{SB}). The accuracy of the
half quantization depends on how the Berry curvature localizes. It follows
from Eq.(\ref{BerryA}) that%
\begin{equation}
\lim_{\Delta \rightarrow 0}\Omega \propto \delta \left( k^{\prime }\right) .
\end{equation}%
Namely, if the Dirac gap $\Delta $ is tiny, the Berry curvature is sharply
localized to the shifted Dirac point $(k_{x}^{\prime },k_{y}^{\prime })=0$,
and the Hall conductivity is almost half quantized as in Eq.(\ref{HalfQ}).
On the other hand, if the Dirac gap is large, the Berry curvature widely
spreads over the Brillouin zone, and the Hall conductivity is not
half quantized as in Eq.(\ref{SB}).

We have derived the formula (\ref{SB}) perturbatively. 
We will check numerically the half-quantization formula Eq.(\ref{HalfQ}) with
Eq.(\ref{sDirac2}) without using perturbation based on the tight-binding model
in the following section.

\section{Tight-binding Model}

The tight-binding representation is constructed by replacing $k_{j}\mapsto
\sin ak_{j}$ and $k_{j}^{2}\mapsto 2\left( 1-\cos ak_{j}\right) $ in the
Hamiltonian (\ref{2dHamil}), where $j=x,y$ and $a$ is the lattice constant.
The Hamiltonian is defined on the square lattice for the $p$-wave magnet,
the $d$-wave magnet and the $g$-wave magnet, while it is defined on the
triangular lattice for the $f$-wave magnet and the $i$-wave magnet.

The tight-binding Hamiltonian corresponding to Eq.(\ref{2dHamil}) is given by%
\begin{equation}
H=H_{\text{Kine}}+H_{\text{R}}+H_{X}+\mathbf{B}\cdot \mathbf{\sigma }.
\label{T-Hamil}
\end{equation}%
The kinetic term and the Rashba term are%
\begin{align}
H_{\text{Kine,Sq}}=& \frac{\hbar ^{2}}{ma^{2}}\left( 2-\cos ak_{x}-\cos
ak_{y}\right) , \\
H_{\text{R,Sq}}=& \lambda \left( \sigma _{y}\sin k_{x}-\sigma _{x}\sin
k_{y}\right)
\end{align}%
on the square lattice. We use them for the $p$-wave magnet, the $d$-wave
altermagnet and the $g$-wave altermagnet.

On the other hand, they are%
\begin{align}
H_{\text{Kine,Tri}}=& \frac{\hbar ^{2}}{ma^{2}}\left( 3-\cos
ak_{x}-\sum_{\pm }\cos a\frac{k_{x}\pm \sqrt{3}k_{y}}{2}\right) , \\
H_{\text{R,Tri}}=& \lambda \sigma _{y}\frac{2}{3}\left( \sin ak_{x}+\sin 
\frac{ak_{x}}{2}\cos \frac{\sqrt{3}ak_{y}}{2}\right)  \notag \\
& -\lambda \frac{2\cos \frac{ak_{x}}{2}\sin \frac{\sqrt{3}ak_{y}}{2}}{\sqrt{3%
}}\sigma _{x}
\end{align}%
on the triangular lattice. We use them for the $f$-wave magnet and the $i$%
-wave altermagnet.

The $X$-wave terms $H_{X}$ are given by

\begin{equation}
H_{p}=J\sigma _{z}\sin ak_{x}
\end{equation}%
for the $p$-wave magnet\cite{pwave,Okumura,EzawaPNeel,He}, 
\begin{equation}
H_{d}=J\sigma _{z}\sin ak_{x}\sin ak_{y}
\end{equation}%
for the $d$-wave magnet\cite%
{SmejRev,SmejX,SmejX2,Zu2023,Gho,Li2023,EzawaAlter,EzawaMetricC,EzawaVolta},%
\begin{equation}
H_{f}=4J\sigma _{z}\sin ak_{x}\sin \frac{ak_{x}+\sqrt{3}ak_{y}}{2}\sin \frac{%
-ak_{x}+\sqrt{3}ak_{y}}{2}
\end{equation}%
for the $f$-wave magnet\cite{GI}, 
\begin{equation}
H_{g}=2J\sigma _{z}\sin ak_{x}\sin ak_{y}\left( \cos ak_{y}-\cos
ak_{z}\right) ,
\end{equation}%
for the $g$-wave magnet\cite{GI}, and%
\begin{align}
H_{i}=& -\frac{16}{3\sqrt{3}}J\sigma _{z}  \notag \\
& \times \sin ak_{x}\sin \frac{ak_{x}+\sqrt{3}ak_{y}}{2}\sin \frac{-ak_{x}+%
\sqrt{3}ak_{y}}{2}  \notag \\
& \times \sin \sqrt{3}ak_{y}\sin \frac{3ak_{x}+\sqrt{3}ak_{y}}{2}\sin \frac{%
-3ak_{x}+\sqrt{3}ak_{y}}{2}
\end{align}%
for the $i$-wave magnet\cite{GI}.

\begin{figure}[t]
\centerline{\includegraphics[width=0.48\textwidth]{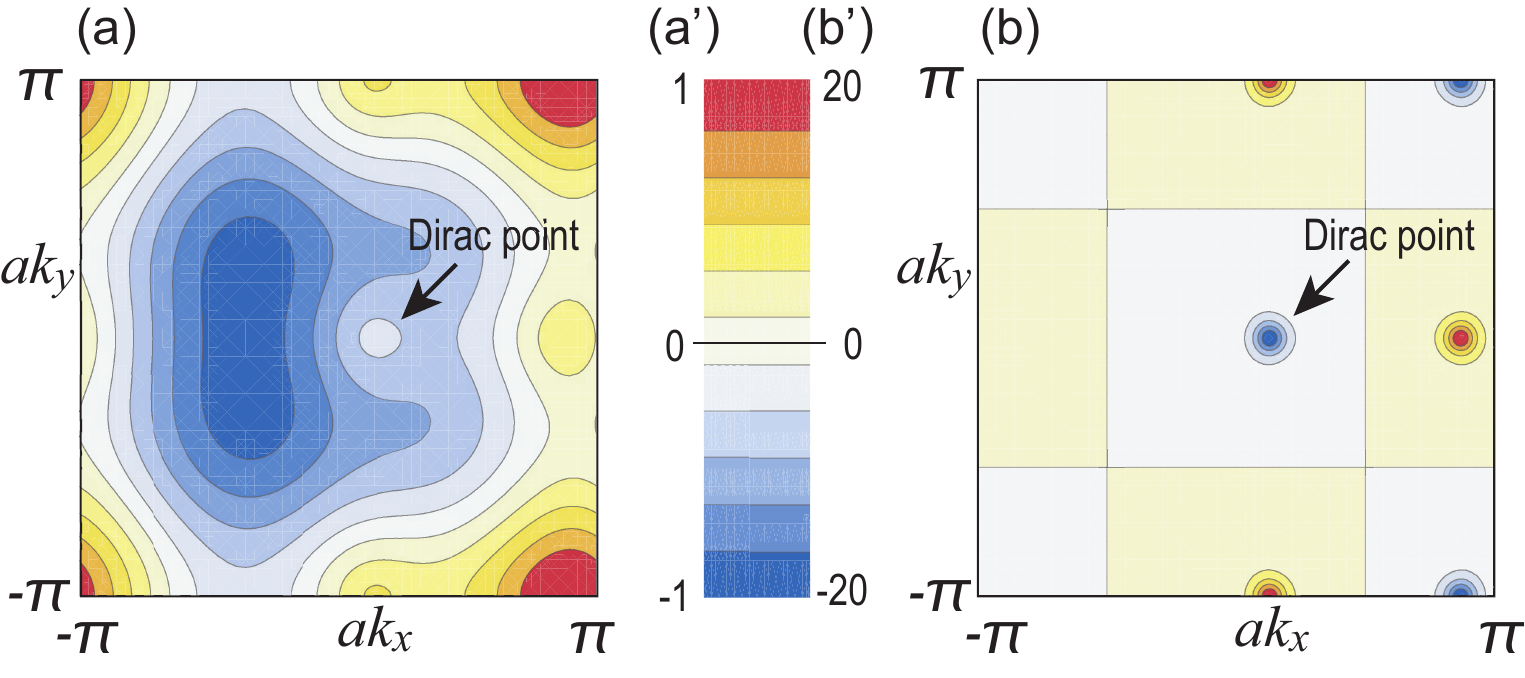}}
\caption{$p$-wave magnet. (a) Contour plot of the energy of the bottom band.
(b) Berry curvature for the bottom band.  (a',b') Color palette for (a,b). 
 We have chosen $\Phi =\protect\pi /2$. 
We have set $Jak_{0}=\protect\varepsilon _{0}/2$, 
$\protect\lambda k_{0}=\protect\varepsilon _{0}$ 
and $\hbar ^{2}k_{0}^{2}/\left( 2m\right) =\protect\varepsilon _{0}/2$.}
\label{FigPBerry}
\end{figure}

\begin{figure}[t]
\centerline{\includegraphics[width=0.48\textwidth]{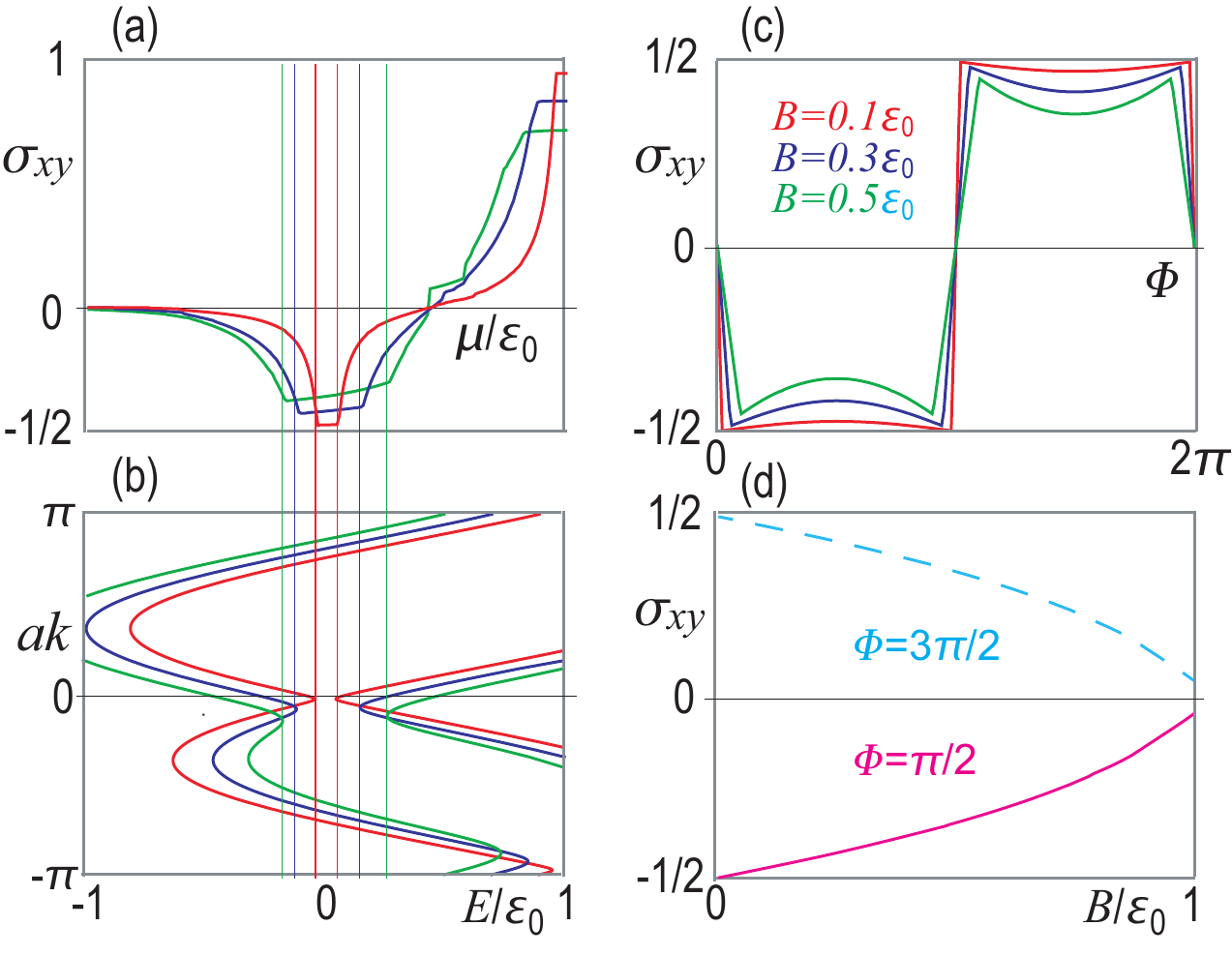}}
\caption{$p$-wave magnet. (a) Hall conductivity  in units of $e^{2}/h$ 
as a function of chemical potential $\protect\mu $. (b) Energy spectrum 
 for $-\protect\pi <ak_{x}<\protect\pi $ and $k_{y}=0.$  (c) Hall
conductivity as a function of angle $\Phi $. In these figures, red curves
indicate $B=0.1\protect\varepsilon _{0}$, blue curves indicate $B=0.3\protect%
\varepsilon _{0}$, and green curves indicate $B=0.5\protect\varepsilon _{0}$%
. (d) Hall conductivity as a function of magnetic field $B$. A magenta curve
indicates $\Phi =\protect\pi /2$ and a cyan curve indicates $\Phi =3\protect%
\pi /2$. We have chosen $\Phi =\protect\pi /2$ in (a), (b), (d). 
See also the caption of Fig.\ref{FigPBerry}.}
\label{FigPphi}
\end{figure}

\subsubsection{p-wave magnets}

We show the contour plot of the energy of the bottom band in the Brillouin
zone in Fig.\ref{FigPBerry}(a), where the positive-energy and
negative-energy regions are made clear  with the color palette given in
Fig.\ref{FigPBerry}(a').  The negative energy part contains one shifted
Dirac point as indicated by an arrow.

The Berry curvature  of each band  is analytically calculated based on
the tight-binding model as $\Omega _{\pm }=\mp \Omega $ with 
\begin{equation}
\Omega =\frac{B_{y}J\lambda \cos k_{x}\cos k_{y}}{2\left( J^{2}\sin
^{2}k_{x}+\left( B_{y}+\lambda \sin k_{x}\right) ^{2}+\left( B_{x}-\lambda
\sin k_{y}\right) ^{2}\right) ^{3/2}}.  \label{p-Berry}
\end{equation}%
We show the Berry curvature $\Omega _{-}$ of the bottom band in Fig.\ref%
{FigPBerry}(b),  where it is found to be sharply localized in the
vicinity of the shifted Dirac point.

We calculate numerically the Hall conductivity (\ref{Hall}) by integrating
the Berry curvatures $\Omega _{\pm }$  below the chemical potential.
The result is shown as a function of the chemical potential $\mu $\ in Fig.%
\ref{FigPphi}(a), where we have set $\Phi =\pi /2$. There is a plateau where
the Hall conductivity is almost half quantized, when the chemical potential
is within the Dirac gap, as is found in the energy spectrum given in Fig.\ref%
{FigPphi}(b).

We show the Hall conductivity as a function of the angle $\Phi $ in Fig.\ref%
{FigPphi}(c), where we have set the chemical potential to the Dirac energy, $%
\mu =E_{\text{Dirac}}$, given by 
\begin{equation}
E_{\text{Dirac}}\equiv \frac{\hbar ^{2}B^{2}}{2m\lambda ^{2}}.
\label{DiracEne}
\end{equation}%
The Hall conductivity is almost half quantized $\sigma _{xy}=-\frac{e^{2}}{h}%
\frac{1}{2}$\ for $0<\Phi <\pi $\ and $\sigma _{xy}=\frac{e^{2}}{h}\frac{1}{2%
}$\ for $\pi <\Phi <2\pi $. It well reproduces the angle dependence of the
Hall conductivity (\ref{pHall}).

We also show the Hall conductivity as a function of magnetic field $B$ in
Fig.\ref{FigPphi}(d), where a magenta curve describes it for $\Phi =\pi /2$,
and a dashed cyan curve describes it for $\Phi =3\pi /2$. It is around $\pm
1/2$ for small $B$.

\begin{figure}[t]
\centerline{\includegraphics[width=0.48\textwidth]{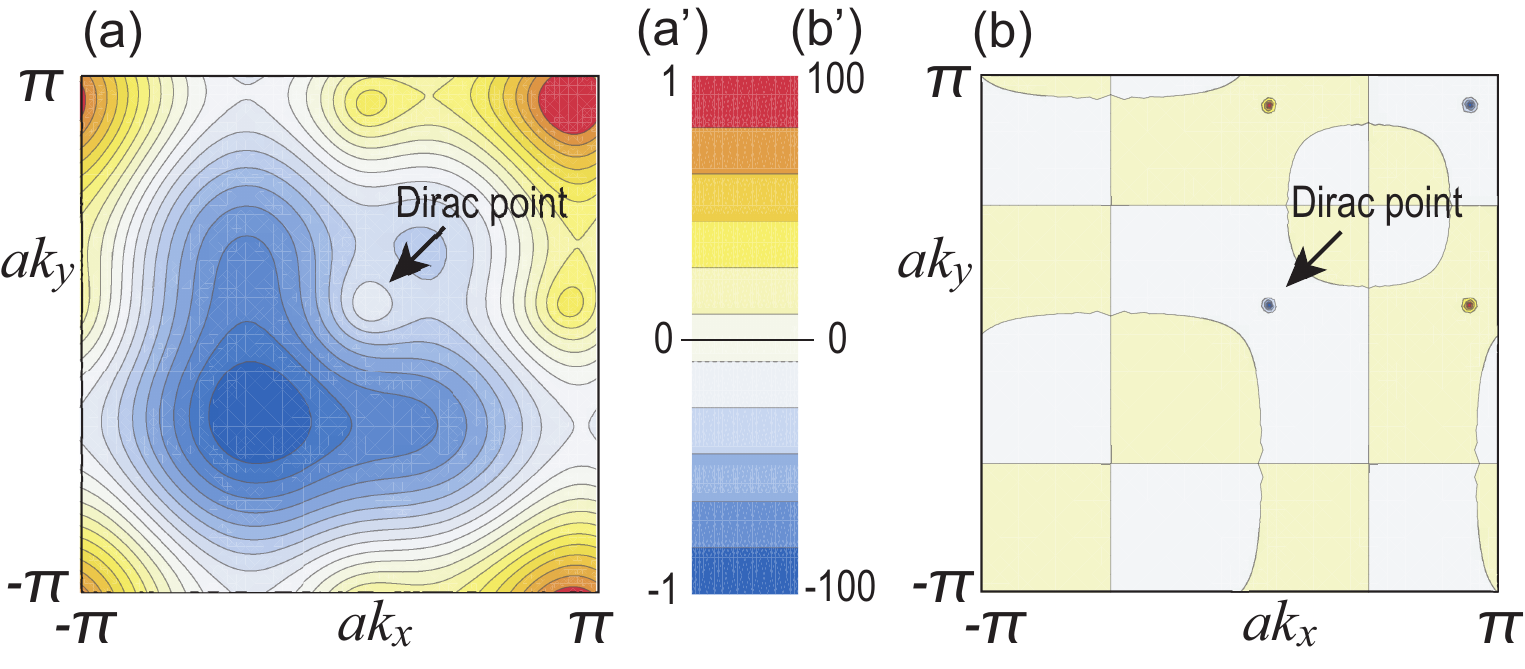}}
\caption{$d$-wave altermagnet. (a) Contour plot of the energy of the bottom
band.  (b) Berry curvature for the bottom band.  (a',b') Color palette
for (a,b).  We have chosen $\Phi =\protect\pi /4$. 
We have set $2Ja^{2}k_{0}^{2}=\protect\varepsilon _{0}/2$, 
$\protect\lambda k_{0}=\protect\varepsilon _{0}$ 
and $\hbar ^{2}k_{0}^{2}/\left( 2m\right) =\protect\varepsilon _{0}/2$.
}
\label{FigDBerry}
\end{figure}
\begin{figure}[t]
\centerline{\includegraphics[width=0.48\textwidth]{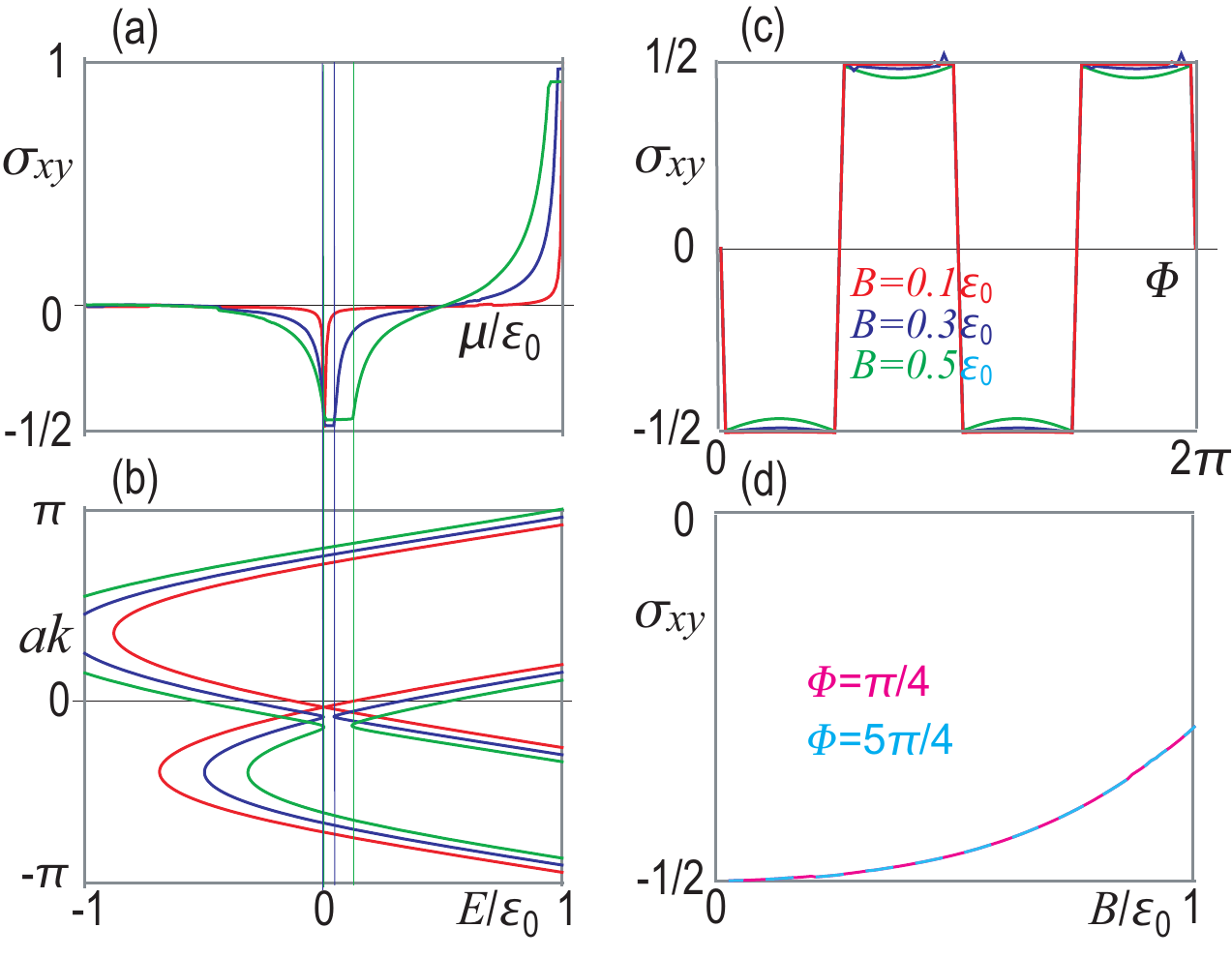}}
\caption{$d$-wave altermagnet. (a) Hall conductivity  in units of $%
e^{2}/h $  as a function of chemical potential $\protect\mu $. (b) Energy
spectrum  for $\left( k_{x},k_{y}\right) =k\left( \cos \frac{\protect\pi 
}{4},\sin \frac{\protect\pi }{4}\right) $ with $-\protect\pi <ak<\protect\pi $%
.  (c) Hall conductivity as a function of angle $\Phi $. (d) Hall
conductivity as a function of magnetic field $B$. A magenta curve indicates $%
\Phi =\protect\pi /4$ and a cyan curve indicates $\Phi =5\protect\pi /4$. We
have chosen $\protect\phi =\protect\pi /4$ in (a), (b), (d). 
We have set $2Ja^{2}k_{0}^{2}=\protect\varepsilon _{0}/2$. See also the caption of Fig.\protect\ref{FigDBerry}.
}
\label{FigDphi}
\end{figure}

\subsubsection{d-wave altermagnets}

The band structure is shown in Fig.\ref{FigDBerry}(a). The Berry curvature
is sharply localized as in Fig.\ref{FigDBerry}(b). The Hall conductivity is
numerically calculated as a function of the chemical potential $\mu $ in Fig.%
\ref{FigDphi}(a), while the energy spectrum is shown in Fig.\ref{FigDphi}%
(b). There is a plateau where the Hall conductivity is almost half quantized.

We show the Hall conductivity as a function of the angle $\Phi $ in Fig.\ref%
{FigDphi}(c). It is almost half quantized, $\sigma _{xy}=-\frac{e^{2}}{h}%
\frac{1}{2}$\ for $0<\Phi <\pi /2$\ and $\pi <\Phi <3\pi /2$. On the other
hand,\ $\sigma _{xy}=\frac{e^{2}}{h}\frac{1}{2}$\ for $\pi /2<\Phi <\pi $
and $3\pi /2<\Phi <2\pi $. It well reproduces the angle dependence in the
formula (\ref{pHall}).

We also show the Hall conductivity as a function of magnetic field $B$ in
Fig.\ref{FigDphi}(d), where a magenta curve describes it for $\Phi =\pi /4$,
and a dashed cyan curve describes it for $\Phi =5\pi /4$. It is around $\pm
1/2$ for small $B$. 

\begin{figure}[t]
\centerline{\includegraphics[width=0.48\textwidth]{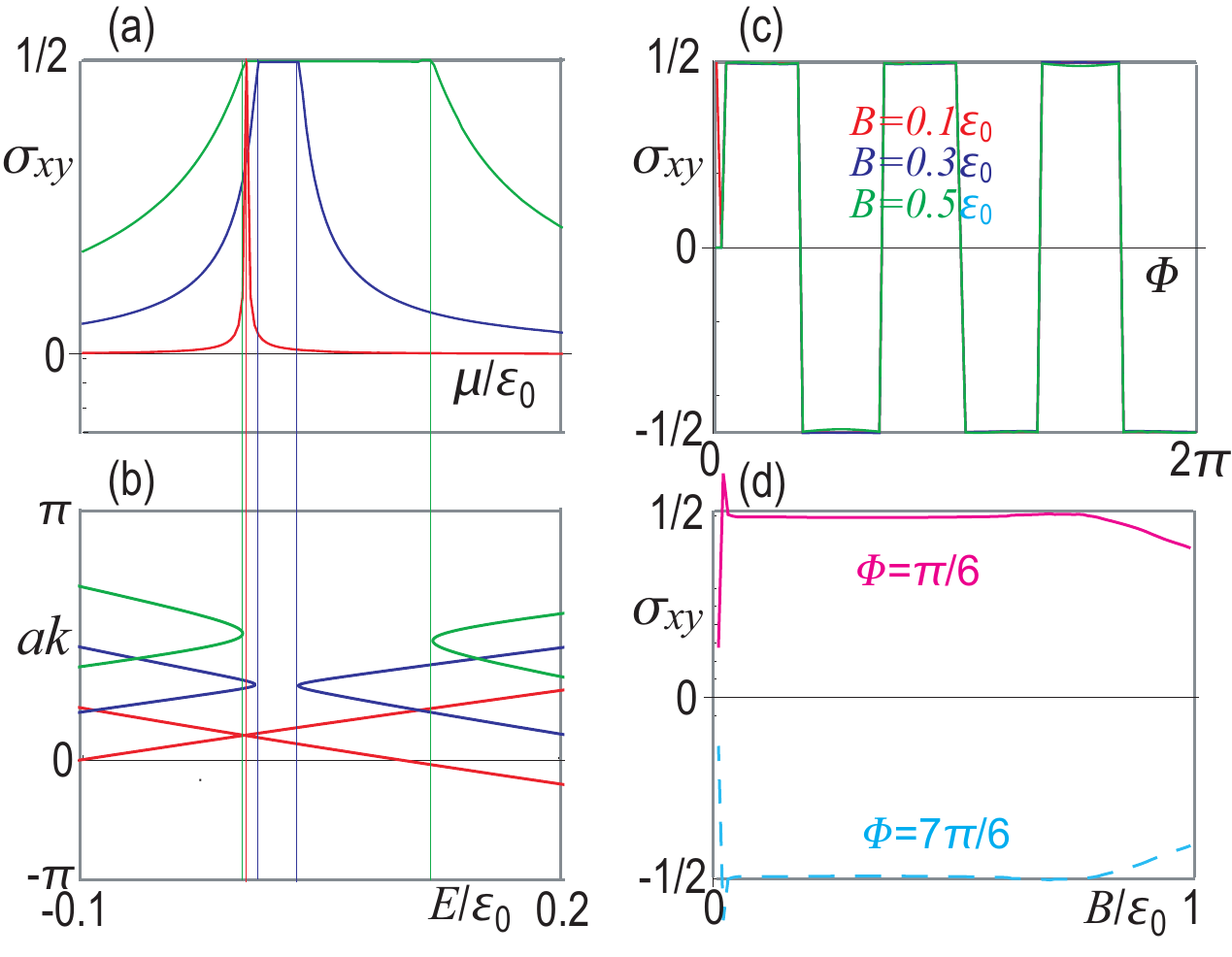}}
\caption{$f$-wave magnet. (a) Hall conductivity  in units of $e^{2}/h$  %
as a function of chemical potential $\protect\mu $. (b) Energy spectrum  %
for $\left( k_{x},k_{y}\right) =k\left( \cos \frac{\protect\pi }{6},\sin 
\frac{\protect\pi }{6}\right) $ with $-\protect\pi <ak<\protect\pi $.  (c)
Hall conductivity as a function of angle $\Phi $. (d) Hall conductivity as a
function of magnetic field $B$. A magenta curve indicates $\Phi =\protect\pi %
/6$ and a cyan curve indicates $\Phi =7\protect\pi /6$. We have chosen $\Phi =%
\protect\pi /6$. We have set $Ja^{3}k_{0}^{3}=\protect\varepsilon _{0}/2$, 
$\protect\lambda k_{0}=\protect\varepsilon _{0}$ 
and $\hbar ^{2}k_{0}^{2}/\left( 2m\right) =\protect\varepsilon _{0}/2$.}
\label{FigFphi}
\end{figure}

\subsubsection{f-wave magnets, g-wave altermagnets and i-wave magnets}

\begin{figure}[t]
\centerline{\includegraphics[width=0.48\textwidth]{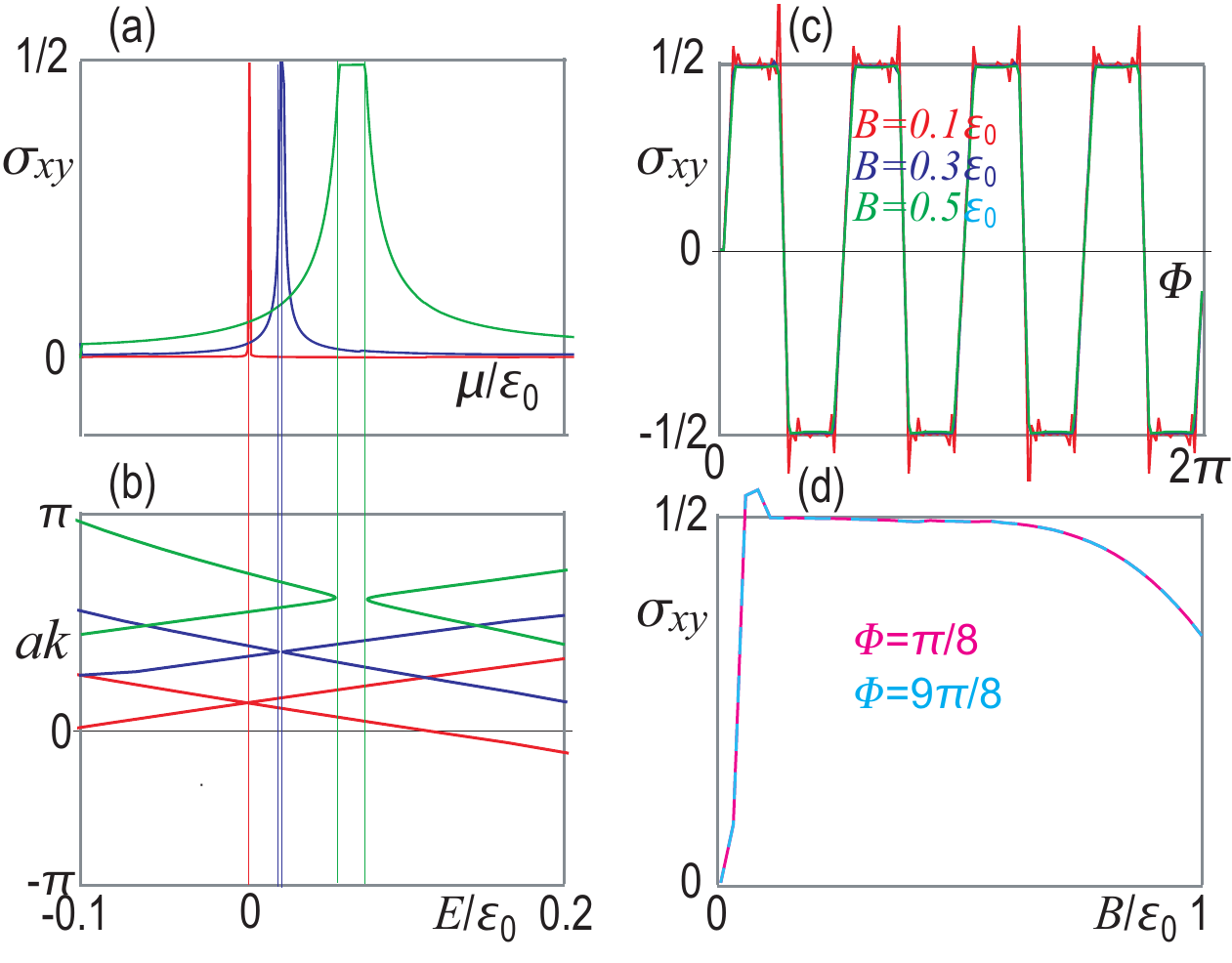}}
\caption{$g$-wave altermagnet. (a) Hall conductivity  in units of $%
e^{2}/h $ as a function of chemical potential $\protect\mu $. We have
chosen $\Phi =\protect\pi /8$. (b) Energy spectrum  for $\left(
k_{x},k_{y}\right) =k\left( \cos \frac{\protect\pi }{8},\sin \frac{\protect%
\pi }{8}\right) $ with $-\protect\pi <ak<\protect\pi $. (c) Hall
conductivity as a function of angle $\Phi $. (d) Magnetic field $B$
dependence of the Hall conductivity. A magenta curve indicates $\Phi =%
\protect\pi /8$ and a cyan curve indicates $\Phi =9\protect\pi /8$. We have
set $4Ja^{4}k_{0}^{4}=\protect\varepsilon _{0}/2$, 
$\protect\lambda k_{0}=\protect\varepsilon _{0}$ 
and $\hbar ^{2}k_{0}^{2}/\left( 2m\right) =\protect\varepsilon _{0}/2$. }
\label{FigGphi}
\end{figure}

\begin{figure}[t]
\centerline{\includegraphics[width=0.48\textwidth]{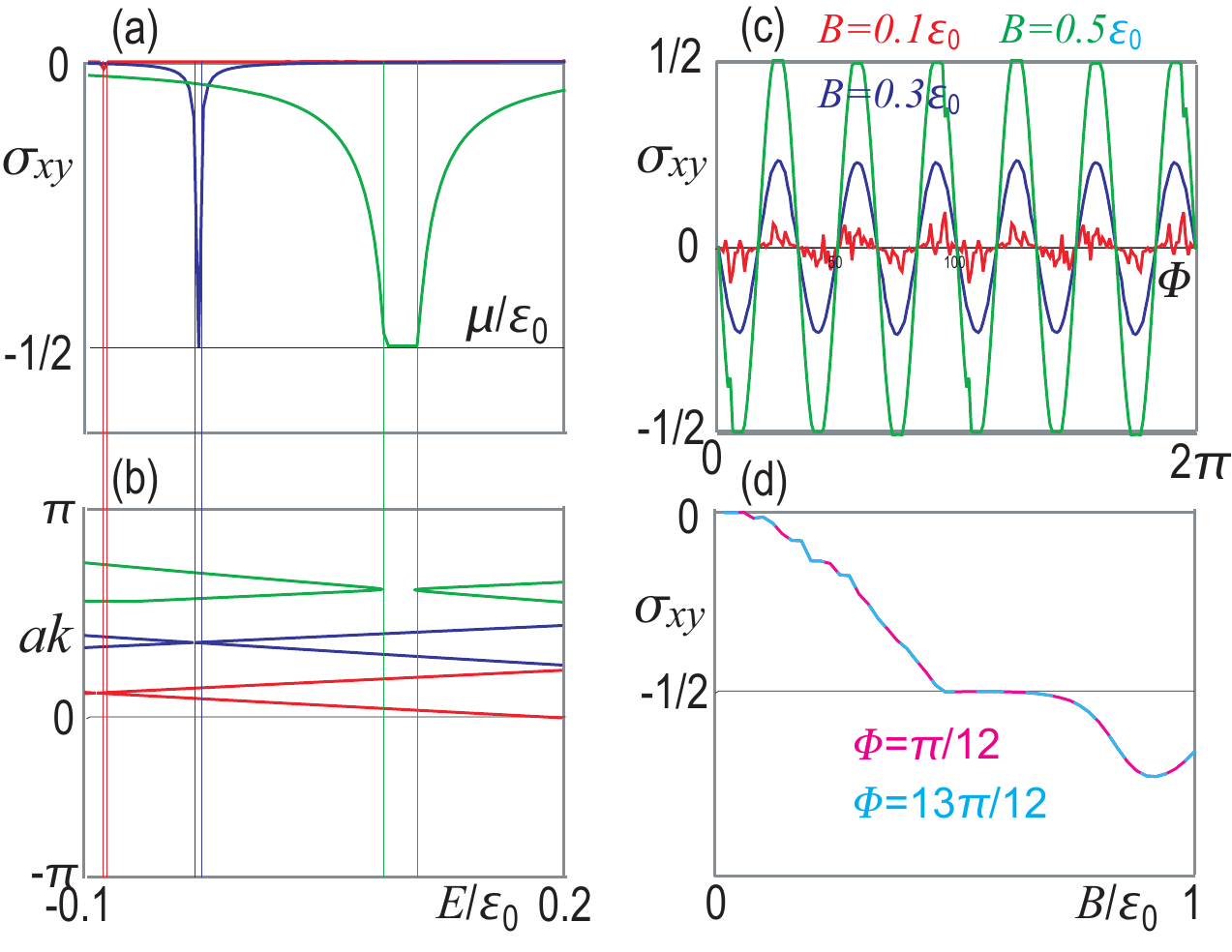}}
\caption{$i$-wave altermagnet. (a) Hall conductivity  in units of $%
e^{2}/h $  as a function of chemical potential $\protect\mu $. (b) Energy
spectrum  for $\left( k_{x},k_{y}\right) =k\left( \cos \frac{\protect\pi 
}{12},\sin \frac{\protect\pi }{12}\right) $ with $-\protect\pi <ak<\protect%
\pi $.  (c) Hall conductivity depends on angle $\Phi $. (d) Hall
conductivity as a function of magnetic field $B$. A magenta curve indicates $%
\Phi =\protect\pi /12$ and a cyan curve indicates $\Phi =13\protect\pi /12$.
We have chosen $\Phi =\protect\pi /12$ for (a), (b) and (d). We have set $%
2Ja^{6}k_{0}^{6}=\protect\varepsilon _{0}/2$, 
$\protect\lambda k_{0}=\protect\varepsilon _{0}$ 
and $\hbar ^{2}k_{0}^{2}/\left( 2m\right) =\protect\varepsilon _{0}/2$. }
\label{FigIphi}
\end{figure}

The Hall conductivity is numerically calculated and shown in Fig.\ref%
{FigFphi} for the $f$-wave magnet, in Fig.\ref{FigGphi} for the $g$-wave
altermagnet and in Fig.\ref{FigIphi} for the $i$-altermagnt, respectively.
They well reproduce the formula (\ref{pHall}). 

For the $g$-wave altermagnets, the Hall conductivity is zero $B=0$. It is
around $\pm 1/2$ for small $B$. 

For the $i$-wave altermagnets, the Hall conductivity is around zero for
small $B$. It contradicts to the half quantization in Eq.(\ref{pHall}). It
would be an artifact in numerical calculations because the Dirac gap is
proportional to $B^{6}$. It is too small for small $B$, where\ the Dirac
energy (\ref{DiracEne}) is away from the gap in the tight-binding model.

\section{Finite temperature effect}

 We show the Hall conductivity as a function of temperature in Fig.\ref%
{FigFiniteT} for the $p$-wave magnet in (a) and for the $d$-wave altermagnet
in (b). Its magnitude decreases as the increase of temperature and leaves
the half-quantized value.

\section{Disorder effect}

The Chern number is rewritten as\cite{Ishikawa,Volovik,Wang,Wang12,Gurarie}%
\begin{equation}
\mathcal{C}=\left( 2\pi \right) ^{-2}\int d^{2}k\int_{-\infty }^{\infty
}d\omega \,\Omega ,  \label{ChernGamma}
\end{equation}%
where the Berry curvature is rewritten as%
\begin{equation}
\Omega =\frac{1}{6}\varepsilon _{\mu \nu \rho }\text{Tr}[G\partial _{\mu
}G^{-1}G\partial _{\nu }G^{-1}G\partial _{\rho }G^{-1}]
\end{equation}%
in terms of the Green function%
\begin{equation}
G\left( \mathbf{k}\right) =[i\omega -H\left( \mathbf{k}\right) -i\Gamma
]^{-1},
\end{equation}%
where $k_{\mu }$, $k_{\nu }$ and $k_{\rho }$ run through $k_{0}\equiv
i\omega $, $k_{x}$ and $k_{y},$ with $i\omega $ referring to the Matsubara
frequency ($\omega $: real) and $\Gamma $ is the self energy representing
the effect of disorder or interactions. $\Gamma $ is renormalized by
shifting the frequency as $\omega ^{\prime }=\omega +\Gamma $. Then, the
Berry curvature is irrelevant to $\Gamma $ because the integral of $\omega $
is taken $-\infty <\omega <\infty $.

\begin{figure}[t]
\centerline{\includegraphics[width=0.48\textwidth]{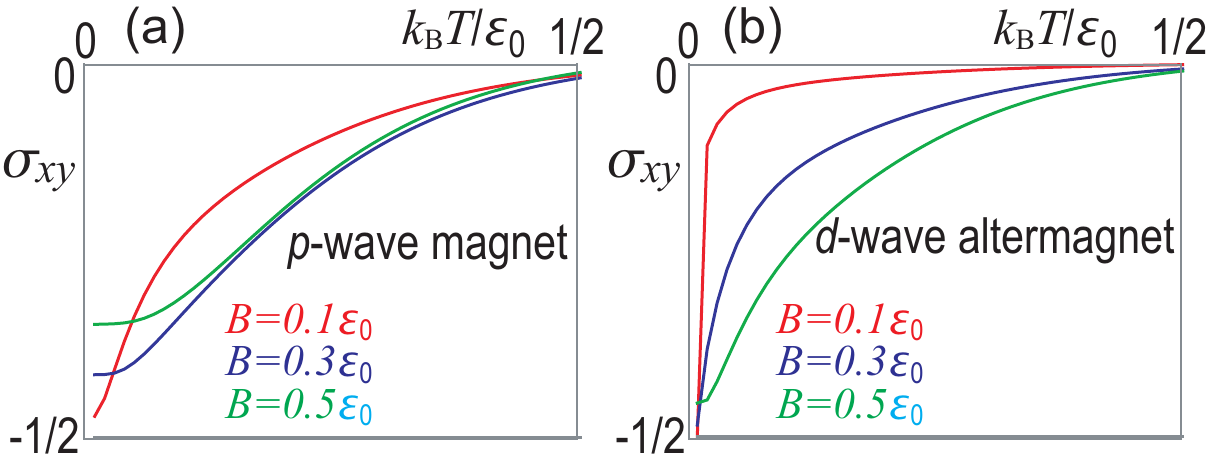}}
\caption{ Finite temperature effect of the Hall conductivity for various
values of $B$. The Hall conductivity in units of $e^{2}/h$ as a function of
temperature $k_{\text{B}}T/\protect\varepsilon _{0}$. We have set $\protect\lambda k_{0}=\protect\varepsilon _{0}$ 
and $\hbar ^{2}k_{0}^{2}/\left( 2m\right) =\protect\varepsilon _{0}/2$. (a) $p$-wave magnet. We have set $Jak_{0}=\protect\varepsilon _{0}/2$. (b) $d$-wave altermagnet. We have set $2Ja^{2}k_{0}^{2}=\protect\varepsilon _{0}/2$. }
\label{FigFiniteT}
\end{figure}

\section{Discussions}

Studying planar Hall effects in the $X$-wave magnets, we have revealed a
universal feature common to all of them. When the Dirac gap is tiny, the
Hall conductivity induced by in-plane magnetic field is almost half
quantized as in Eq.(\ref{pHall}). This formula implies that  the
magnitude of the Hall conductivity is proportional to sgn$\left( J\sin
N_{X}\Phi \right) $. It is possible to determine $N_{X}$ by measuring the $%
\Phi $ dependence. On the other hand, the Dirac gap is given by $\Delta
\propto B^{N_{X}}$. Then, the fine tuning of the chemical potential is
necessary to adjust the chemical potential within the Dirac gap for large $%
N_{X}$ and for small $B$.

The absolute value of the Hall conductivity takes the maximum value at $\Phi
=\left( \pm \pi /2+2n\pi \right) /N_{X}$\ with $n=0,\cdots ,N_{x}-1$ in the $%
X$-wave magnet. In particular, we have studied the Hall conductivity as a
function of magnetic field $B$\ by fixing $\Phi =\pi /\left( 2N_{X}\right) $%
\ and $\pi /\left( 2N_{X}\right) +\pi $ in Figs.3(c), 5(c), 6(c), 7(c) an
8(c) by the following reason. External magnetic field is introduced by
applying current to a solenoid. By changing the direction of the applied
current continuously, the angle $\Phi =\pi /\left( 2N_{X}\right) $\ is
continuously flipped to $\Phi =\pi /\left( 2N_{X}\right) +\pi $. The Hall
conductivity is odd for $\Phi =\pi /\left( 2N_{X}\right) $\ and $\pi /\left(
2N_{X}\right) +\pi $\ for the $p$-wave magnet and the $f$-wave magnet. On
the other hand, it is even for the $d$-wave altermagnet, the $g$-wave
altermagnet and the $i$-wave magnet.

 The $d$-wave, $g$-wave and $i$-wave altermagnets have collinear
antiferromagnetic spin texture, while $p$-wave magnets have helical spin
texture. By applying external magnetic field, the spin texture is canted to
the direction of magnetic field. It produces an additional term $\mathbf{M}%
\cdot \mathbf{\sigma }$, where $\mathbf{M}$\ is the additional magnetization
induced by magnetic field. However, it is renormalized by setting new
magnetic field $\mathbf{B}^{\prime }=\mathbf{B}+\mathbf{M}$. Hence, the
effect of the magnetization can be included in the magnetic field term as $%
\mathbf{B}^{\prime }\cdot \mathbf{\sigma }$ in the Hamiltonian (\ref{2dHamil}%
).

It was theoretically proposed that CeNiAsO is a $p$-wave magnet\cite{pwave}%
.~It was also theoretically proposed that a $p$-wave magnet is realized in
graphene by introducing spin nematic order\cite{Roy}. Experiments on $p$%
-wave magnets\ were reported\cite{Yamada} in Gd$_{3}$Ru$_{4}$Al$_{12}$\ and
reported\cite{Comin} in NiI$_{2}$. A $d$-wave magnet is realized in RuO$_{2}$%
\cite{Ahn,SmeRuO,Tsch,Fed,Lin}, Mn$_{5}$Si$_{3}$\cite{Leiv} and FeSb$_{2}$%
\cite{Mazin} organic materials\cite{Naka}, perovskite materials\cite%
{NakaB,NakaRev}, and twisted magnetic Van der Waals bilayers\cite{YLiu}. It
was theoretically proposed that an $f$-wave magnet is realized in
rhombohedral trilayer graphene by introducing spin nematic order\cite{Roy}.
A $g$-wave altermagnet and an $i$-wave altermagnet are realized in twisted
magnetic Van der Waals bilayers\cite{YLiu}.

The sgn$J=\pm 1$ of the $d$-wave altermagnet has been proposed to use as an
Ising bit, where the bit is expected to flip very fast\cite%
{SmejX,SmejX2,SmejRev}. In a similar way, we may use the sgn$J$ of other $X$%
-wave magnets as a bit. As in the case of the $d$-wave altermagnet, there is
zero-net magnetization in the $X$-wave magnets, which would provide us with
ultra-fast and ultra-dense nonvolatile memories.

This work is supported by CREST, JST (Grants No. JPMJCR20T2) and
Grants-in-Aid for Scientific Research from MEXT KAKENHI (Grant No. 23H00171).

\end{document}